# VISUALIZING COLLECTIVE DISCURSIVE USER INTERACTIONS IN ONLINE LIFE SCIENCE COMMUNITIES

Dhiraj Murthy, Alexander Gross, and Stephanie Bond

Social Network Innovation Lab
Bowdoin College
7050 College Station
Brunswick, Maine, 04011-8470, USA
e-mail: dmurthy@bowdoin.edu

## ABSTRACT

This paper highlights the rationale for the development of BioViz, a tool to help visualize the existence of collective user interactions in online life science communities. The first community studied has approximately 22,750 unique users and the second has 35,000. Making sense of the number of interactions between actors in these networks in order to discern patterns of collective organization and intelligent behavior is challenging. One of the complications is that forums - our object of interest - can vary in their purpose and remit (e.g. the role of gender in the life sciences to forums of praxis such as one exploring the cell line culturing) and this shapes the structure of the forum organization itself. Our approach took a random sample of 53 forums which were manually analyzed by our research team and interactions between actors were recorded as arcs between nodes. The paper focuses on a discussion of the utility of our approach, but presents some brief results to highlight the forms of knowledge that can be gained in identifying collective group formations. Specifically, we found that by using a matrix-based visualization approach, we were able to see patterns of collective behavior which we believe is valuable both to the study of collective intelligence and the design of virtual organizations.

## INTRODUCTION

Cyberinfrastructure is playing an increasingly important role in organizations. In virtual organizations, this is even more true. However, many of the decisions made in designing spaces for virtual interactions are not based on data derived from models of collective intelligence. Furthermore, there is an immense store of knowledge in terms of decision analysis, problem solving, and the development of trust when applied to understanding data from online interactions.

In this paper, we describe our study of two global virtual life science communities which facilitate the creation of virtual organizations, problem solving of scientific questions, and interactions between junior and senior faculty/scientists amongst other things. Our work has been most interested in the ability of these communities to facilitate the construction of virtual organizations (VOs). In our study of this, we have needed to examine large amounts of data which reflects the collective interactions of users. In terms of variables, we have been particularly interested in measuring how trust forms in these communities (where individuals have most often not met face-to-face).

This paper explores our approach to investigating one of the research questions of this larger project : how can we more intelligently understand collective formations of trust and sentiment in these communities. To explore this question, we studied the message forums within these communities. We sampled 53 forums with 1292 unique users who interacted 5823 times. The total user population which our sample draws from is approximately 57,750. We studied each post within the selected forums and specified which users were interacting (i.e. ego interacts with which node(s)), the frequency of the interactions in the forum, whether the interactions were fostering trust, and what sentiment the interactions were fostering.

We took the data from the collective set of data spanning both communities and created a matrix-based visualization tool, BioViz, in order to better understand the collective formations that were emerging. Although there are various software packages for visualization (including robust social network analysis software), we saw the potential for the development of custom lightweight tools that allow for quick exploration, identification, and confirmation of collective formations within these communities. We also felt that developing an easy to understand visualization structure to display this data would better enable lay consumers to learn from our work on collective intelligence.

**BACKGROUND**

The concept of a map of interactions is important to the study of social networks as collectivities of users. A map of interactions is used widely in the life sciences (e.g. (Giot, et al., 2003)) and can be loosely defined as any documentation of interactions between a set of entities that constitute a network, though the definition of the *interactions* and *entities* for a given network is open for interpretation. This openness allows network researchers great freedom to create networks maps that explore many different types of interactions, as well as different understandings of what constitutes the set of entities. This richness guarantees that for any network there may be a near infinite number of ways to create a map over that network, each carrying the potential to illuminate different aspects and characteristics of the networks as a whole. Typical examples familiar in the realm of social computing include: mapping users to users with friendship relationships on the Facebook social network (Lewis, Kaufman, Gonzalez, Wimmer, & Christakis, 2008), or mapping users to users based on a following relationship in the Twitter micro-blogging service. Other examples might include mapping websites to websites they link to (McNally, 2005), mapping individuals to papers they are authors on (Newman, 2001), or mapping users to online forums to which they have contributed.

**Types of visual network maps**

The collective data that constitutes a map of interactions of a virtual community may be expressed in many forms, from a raw text based dataset (e.g. a Pajek .net file) to complex layered visualizations. In addition to a set of defined entities and interactions, the data set may also often include important metadata about each entity and interaction. This metadata can be used to organize the data in different ways, create subsets or partitions over the data, or in the case of visualization can be used to map to visual properties that represent the members of the network and their interactions. See Figure 1 as an example.

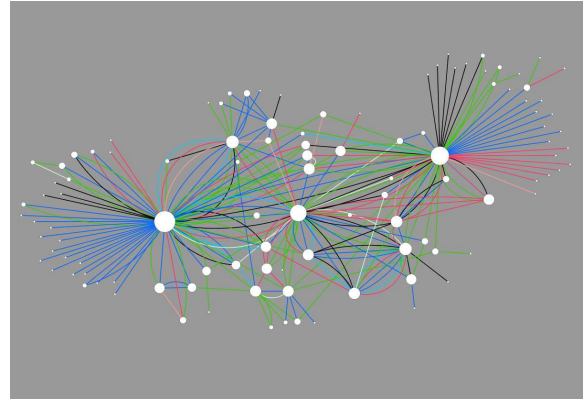

*Figure 1: A node-based network map of the communications between users in one life science community. Interactions are color-coded for sentiment, while user nodes are size-coded for total number of posts in the forum.*

*Node-based Network Maps*

A common way to visualize networks of users is to render a set of nodes representing entities with lines drawn between these nodes representing their interactions. Metadata will often be mapped to visual attributes like size or color of node and lines. The example given in (Figure 1) is typical of interaction maps of this class. Often these types of visualizations can make for a good illustration, but beyond a certain threshold of entities and interactions the space of the visualization can quickly become too cluttered to be able to interpret information carried by the data at all but the highest of levels. Also the placement of nodes often does not have any correlation to the data, the results being that there are an infinite number of ways to distribute nodes in the space, but only a small percentage of these will be conducive toward a broader understanding of the data. Although many algorithms have been developed for this purpose, there is no definitively optimal strategy for the visual distribution of nodes in a visualization space. For small networks, this is not a problem as they are able to be represented well in the visual field. But, for applications including large-scale social computing, crowd sourcing, and collective behavior, the multitude of nodes and job creating visualizations which are extremely dense. Figure 2 shows a network map of the Internet created in 2005 by the Opte Project (The Opte Project [CC-BY-2.5] via Wikimedia Commons). This rendering has been highly optimized for legibility using both automated node layout algorithm as well as human assisted post-processing to be as communicative as possible but with being able to get into the data and explore it in depth, the most it conveys at-a-glance is perhaps its distinctive branching structure as well as perhaps the relative distribution of top-level domains which are color-coded within the network map.

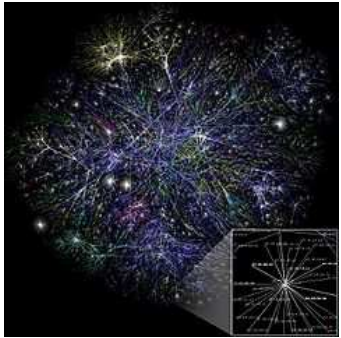

Figure 2: Map of the internet created by The Opte Project (The Opte Project [CC-BY-2.5] via Wikimedia Commons)

*Matrix-based Network Maps*

As a visual design construct, a matrix is a familiar pattern because of its common use in everyday data presentation like simple spreadsheets. A matrix- or table-based design pattern features two sets of entities. The matrix itself represents the Cartesian product of these two sets, with one box in the matrix to correspond with each possible pairing of a member of the first set to a member of the second. In the context of a network map, the two sets are often the same: the set of users under consideration. The attributes stored in each box of the matrix are thought of as characterizing the relationship between two entities. This could take the form of a simple true/false dichotomy to indicate existence of a relationship, a numeric value to indicate the weight of some interaction, or further one could imagine a vector of values corresponding to weight of different kinds of interactions thus forming a natural set of layers on the dataset. In the realm of visualization one can also imagine the value stored in the matrix being mapped to a color. This kind matrix-based visualization is sometimes called a heat map because of its similarity to the idea of temperatures being distributed over a pair or latitudinal and longitudinal coordinates on a weather map.

Within the context of user-to-user interactions, matrix-based visualizations have some interesting properties that offer a different set of opportunities for communicating information about a network. A simple 4x4 matrix model of network interactions can be seen in Figure 3.

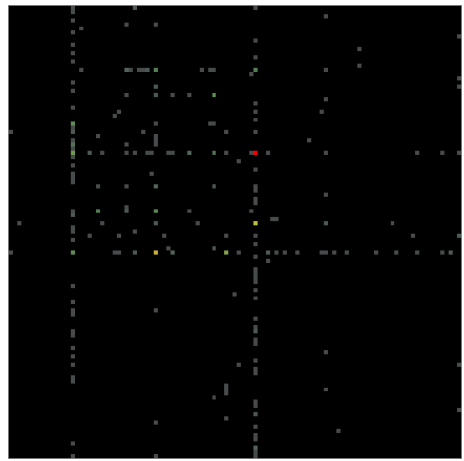

Figure 3: A matrix-based visualization template showing possible interactions between 4 users.

Matrix-based visualization can lead to reduced visual clutter because in practice nodes and line are eliminated leaving just the interactions to be sorted into the appropriate matrix box. Another important feature is the fact that matrix-based visualizations immediately provide more information than the equivalent node-based network map because it represents the complete space of possible interactions including those interaction that haven't yet been formed. In this way matrix based networks maps are often quite sparse, yet at the same time minimal with respect to the data. They can make it easy to see where in the network interactions are happening. Figure 4 shows the matrix-based visualization of the same forum mapped in Figure 1.

Figure 4: A matrix-based visualization of the same forum illustrated in Figure 1.

It is easy to see which users sent a lot of message, and which users received the most messages. It is also easy to pick out which specific user pairs had the highest frequency of interactions. To discover this information from the node-based visualization one would have to spend a lot more time with the image and explore it in detail.

## METHODOLOGICAL APPROACH

In order to explore our data through the lens of matrix-based visualization, we set out to create a lightweight tool for the rendering of such visual matrices from the forum data we had collected. We wanted to develop a visualization system that used a matrix-based design pattern in order to illustrate the intensity of interaction between users in certain topic-based online forums composed of life scientists. It was also important to us for the visualization to be interactive such that one could acquire more detailed information about interactions and quickly compare the network of one forum to another. The purpose of this was to be able to see manifestations of collective intelligence and to be able to make comparative judgments across forums. For example, in a forum about chemical testing methods, are forms of interaction which are both collective and intelligent emerging or are there one or two actors doing all the problem-solving and being the brokers of intelligence.

### Data Model
The general layout of the visual matrix (Figure 3) includes every user posting within a forum across both axes. At the intersection between two users, the values corresponding to variables describing interactions will be recorded. Users are unable to interact with themselves, thus the boxes at these intersections are filled by 'X's.

### Process
First, raw information about the forums, their users, and communication were collected using data-mining techniques approved by the organizations sponsoring the communities. This data was evaluated by our research team and values for variables were all manually recorded. Of particular value, we manually determined which users were interacting together as machine learning processes were not accurate enough to determining interactants. Specifically, we evaluated interactants to determine whether they trusted each other as collaborators, as an information source, or as a confidant. In terms of negative levels of trust, we determined whether they mistrusted the intentions of the other user or mistrusted the other user as an information source. Following trust literature (Butler, 1999), we collapsed these categories in the visualization tools to 'Trust', 'Neutral', and 'Mistrust'. For sentiment, we recorded 'Negative', 'Positive', and 'Unrelated'. Following sentiment literature (Kim & Hovy, 2006), we added a 'Neutral' category. As this is the most important piece of information in order to generate matrix-based visualizations, a high value is placed on accuracy.

Second, we evaluated the content of each post to determine the type and level of trust exhibited in the posting and the overall sentiment of the communication between users. This provided important metadata about each interaction.

## MATRIX-BASED VISUALIZATION TOOL

### Structure
The structure from the raw data manipulation to the visualization is shown below in pipeline format (Figure 5).

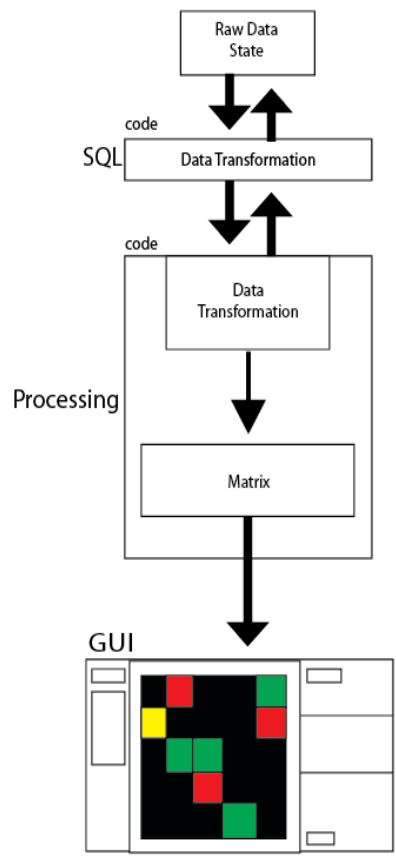

*Figure 5: Diagram of information flow for the matrix-based visualization tool.*

The raw data is gathered through calls to a database using SQL queries. This data is manipulated into the appropriate format to be read by the processing engine. The first call to the database retrieves each forum name and sends it to the processing engine, which in turn will add the forum names to the dropdown list. Once the user chooses a forum name, another call to the database will retrieve the remaining data necessary. The data will be manipulated in order to create the appropriate dimensions and color for each square of the visualization matrix. Once the matrix has been initialized, it will appear in the center section of the Graphical User Interface.

### Interface

The initial storyboard outlines the original features (Figure 6). This design framework was inspired by the functionalities needed and ease of operation.

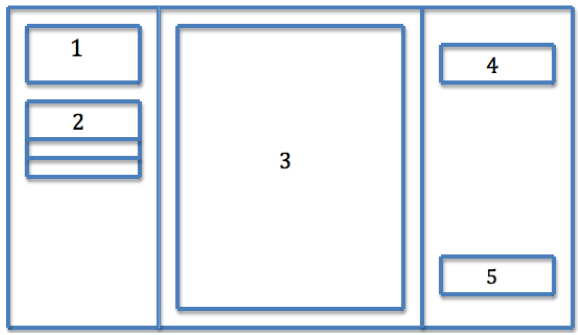

*Figure 6: Outline of interface for a matrix-based visualization tool.*

The tool is separated into three sections that interact with one another, but serve different purposes. The left hand section (see Figure 6) contains the objects that will set the parameters of the visualization. The user interacts with this column (objects 1/2) in order to determine which forum will be visualized. The middle section (object 3) is devoted to the visualization of the data. The right hand section (objects 4/5) contains the keys to explain the visualization as well a feature to save the visualizations once generated.

1. Submit button: This runs the code that fills the graphic window (see object 3).
2. Dropdown list: Contains a complete list of forums from both life science communities.
3. Graphic window: This is where the visualization matrix will be displayed.
4. Color legend: This shows the range of colors that correspond to the frequency of interactions between two users.
5. Save Image button: This triggers a pop up window with image export options.

### PURPOSE

Though the purpose behind the creation of this visualization tool was for our own research and, we believe this model is of value to others studying collective behaviors in large groups of users. We found our visualization tool to be highly successful in helping us determine just how 'collective' the forums actually were. By examining the generated visualizations, we were able to not only develop hypotheses to test using social network analysis (SNA)(Scott, 1988), but we were able to immediately decipher whether the group was collectively intelligent or whether one or two individuals were the source of the 'intelligence'. This is highly valuable to the ends of our project in that virtual organizations can be better designed to facilitate more collective interactions, which can increase trust and sentiment between actors.

Our hope was that through these visualizations we would be able to evaluate which groups were problem-solving, how users were configured and whether they were more collective or individualistic. If collective, were there high levels of trust/positive sentiment across the groups or were high levels of trust/sentiment concentrated amongst just a handful of users? If there was confirmed correlation between identifiable visual phenomena and derived data, then it would be reasonable to assume the visualizations themselves could be evaluated for certain properties as indicators that a network has certain properties, even before these results have been further analyzed using SNA. In this way, the visualization tool could become a research tool to identify points of interested for further study in large complex datasets. The rest of the paper details our experimental finding on the ability and ways in which matrix based visualization can convey communicate important network features visually.

### RESULTS AND CONCLUSIONS

The result of our experimental visualization tool, BioViz, confirm that it is a valuable method for deciphering the ways in which collective intelligence may or may not be manifested in online discursive groups. Interestingly, many of the visualizations revealed forums in which one or two people dominated the discursive space in terms of frequency of interactions, levels of trust, and high levels of positive sentiment. Rather than being a manifestation of collective intelligence, these forums revealed more singular sources of intelligence and problem-solving.

### Symmetry

Figure 7 shows a network map on a small forum that is nearly symmetrical where the axis of symmetry is a diagonal line from the top left to the bottom right corner.

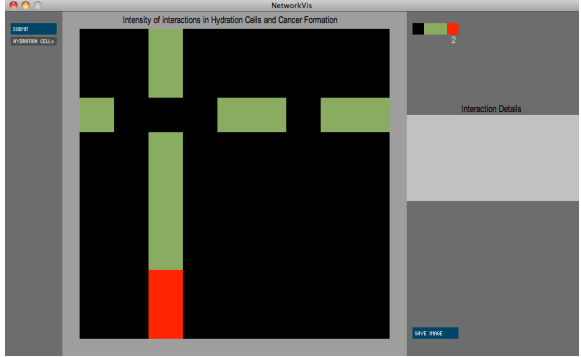

*Figure 7: Example Visualization; Hydration Cells and Cancer Formation forum*

The symmetry gives insight to the nature of interactions within this forum. In this case it means that those who send messages to userX generally receive messages from userX. In this case all users sent at least one message to userX and only three who did not receive a response. We know that symmetry in a matrix-based network map indicates this because both axes of the matrix list users in the same order (See Figure 8)

*Figure 8: Similar users interact in boxes that are symmetrical across the axis of symmetry.*

On a global scale, this suggests that symmetry is highly correlated with responsiveness of a forum. Conversely, visualizations without high levels of symmetry could indicate that a forum could be characterized as a place where posts often go unanswered. In our opinion, this is symptomatic of forums which lack characteristics of collective intelligence.

**Scan Lines**

Consider another observation relating to Figure 7. Instead of looking at symmetry, consider the matrix column-by-column or row-by-row. The existence of highly-defined rows and/or columns may indicate the presence of forum leaders. For instance a strong horizontal line indicates a person who was sent a lot of messages to a lot of different people, this person maybe be facilitating a discussion or answering a lot of questions. Strong vertical lines may indicate a leader or polarizing member of a forum. They are a person whom many have responded to. The may have asked a really interesting question, or said something controversial in order to draw so many responses. A more robust visualization model could allow even more insight by incorporating trust and sentiment data into the visualization. In our opinion, this has implications for the design of discursive organizational spaces.

**Dispersion**

Another attribute of matrix-based visualizations of potentially collective groups is the concept of dispersion. In order to consider dispersion, one must consider the matrix from the top most level. Whether interactions tend to be isolated to a few forum leaders or if interactions are dispersed more evenly throughout the matrix should be considered. When interactions tend to be more concentrated (as seen in Figure 9), this can indicate a forum where there are one on more major conversations taking place with a small set of active forum leaders and another larger groups of users participating rarely to make some small comment on the ongoing discussions. The lines of colored matrix squares demonstrate that individual users are dominating the forum. In this particular example, we have explicit confirmation as this was from a moderated forum on a specific topic where discussions are intentionally focused on a few subjects important to the moderator. An example of a more dispersed forum can be seen in Figure 10. Here, the clustering and definition of interaction is more evenly distributed through the visualization space. This suggests the last of any major topic threads in the forums, but instead a kind of free for all where issues come up and are dispensed with quickly. This is supported by our qualitative analysis. The Cell Lines forum that is being represented in Figure 10 is a space where life science researchers share procedures related to the culturing of cell lines. The forum users are mostly other researchers of a similar background who may provide a quick answer to a question. But, once these highly technical questions are answered there is little need for further discussion. It is cases like that displayed in Figure 10, where we see collective intelligence emerge. Specifically, our qualitative analysis indicated a purpose of problem-solving geared towards the culturing of cell lines. The visualization clearly shows a collective rather than leader-oriented (see Figure 9) problem-solving orientation. When analyzing a large quantity of discursive spaces within two virtual communities, we found this method of visualization highly effective in supporting our identification of forms of collective intelligence within the groups.

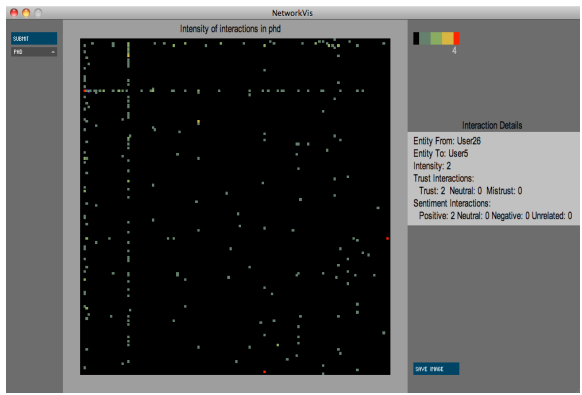

Figure 9: Example Visualization; PHD forum

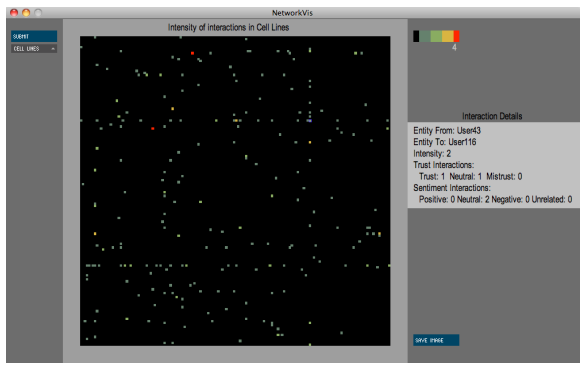

Figure 10: Example Visualization; Cell Lines forum

**ACKNOWLEDGMENTS**

This research was supported by the National Science Foundation (grant number 1025428).

**REFERENCES**

Butler, J. K. (1999). Trust Expectations, Information Sharing, Climate of Trust, and Negotiation Effectiveness and Efficiency. *Group & Organization Management, 24*(2), 217-238.

Giot, L., Bader, J. S., Brouwer, C., Chaudhuri, A., Kuang, B., Li, Y., et al. (2003). A Protein Interaction Map of Drosophila melanogaster. *Science, 302*(5651), 1727-1736.

Kim, S.-M., & Hovy, E. (2006). *Extracting opinions, opinion holders, and topics expressed in online news media text*. Paper presented at the Proceedings of the Workshop on Sentiment and Subjectivity in Text.

Lewis, K., Kaufman, J., Gonzalez, M., Wimmer, A., & Christakis, N. (2008). Tastes, ties, and time: A new social network dataset using Facebook.com. *Social Networks, 30*(4), 330-342.

McNally, R. (2005). Sociomics! Using the IssueCrawler to map, monitor and engage with the global proteomics research network. *PROTEOMICS, 5*(12), 3010-3016.

Newman, M. E. J. (2001). The structure of scientific collaboration networks. *Proceedings of the National Academy of Sciences, 98*(2), 404-409.

Scott, J. (1988). Social Network Analysis. *Sociology, 22*(1), 109-127.

The Opte Project [CC-BY-2.5] via Wikimedia Commons (Producer).